\documentclass[5p,twocolumn,times,number]{elsarticle}

\usepackage{graphicx}
\usepackage{amsmath}

\pdfmapfile{+txfonts.map}

\biboptions{sort&compress}

\def\be{\begin{equation}}
\def\ee{\end{equation}}
\def\beqn{\begin{eqnarray}}
\def\eeqn{\end{eqnarray}}
\def\no{\nonumber}
\def\ba{\begin{array}{c}}
\def\bat{\begin{array}{cc}}
\def\ea{\end{array}}
\def\bi{\begin{itemize}}
\def\ei{\end{itemize}}

\def\cL{{\cal L}}

\def\cO{{\cal O}}

\def\cY{{\cal Y}}

\newcommand{\bel}[1]{\be\label{#1}}

\begin{document}

\begin{frontmatter}

\title{Status after the first LHC run:
Looking for new directions in the physics landscape}

\author{Antonio~Pich}

\address{Departament de F\'\i sica Te\`orica, IFIC, Universitat de Val\`encia -- CSIC, Apt. Correus 22085, E-46071 Val\`encia, Spain}

\begin{abstract}
The LHC data have confirmed the Standard Model as the correct theory at the electroweak scale. It successfully explains the experimental results with high precision and all its ingredients, including the Higgs boson, have been finally verified.
At the same time, the negative searches for signals of new phenomena
challenge our previous theoretical wisdom on new-physics scenarios.
\end{abstract}


\end{frontmatter}


\section{A Higgs-like Boson Discovered}
\label{sec:Higgs}

Combined with all previous experimental tests, the first LHC run has established the Standard Model (SM) as the appropriate description of the electroweak interactions at the energy scales explored so far \cite{Pich:2015tqa}. A scalar with the expected properties of a Higgs boson has been discovered \cite{Aad:2012tfa,Chatrchyan:2012ufa}; it has a spin/parity consistent with the SM $0^+$ assignment and its mass \cite{Aad:2015zhl},
%
\bel{eq:M_H}
M_H = (125.09\pm 0.24)\;\mathrm{GeV}\, ,
\ee
is in the range preferred by global fits to precision electroweak data \cite{Baak:2014ora}. The experimental results are in remarkable good agreement with the theoretical predictions, exhibiting an overwhelming success of the SM paradigm.

The new scalar couples to fermions and gauge bosons ($W^\pm$, $Z$, $\gamma$, $G^a$) with the strengths predicted by the Higgs mechanism \cite{Higgs:1964pj,Englert:1964et,Guralnik:1964eu,Kibble:1967sv}. The measured $H$ production cross section, which is dominated by gluon-fusion ($GG\!\to\! t\bar t\!\to\! H$), confirms the existence of a $t\bar tH$ coupling with the SM size, and
the sign of the top Yukawa has been verified in the decay $H\to\gamma\gamma$ \cite{Khachatryan:2014jba,Aad:2014eha}
through the destructive interference of the $W^+W^-$ and $t\bar t$ loop contributions.
The tree-level decays $H\to W^{\pm *} W^{\mp}, Z^* Z$ directly test the electroweak gauge couplings of the Higgs \cite{Khachatryan:2014jba,Aad:2014eva}. In addition, we have now strong evidence for the $H$ coupling to $b\bar b$ and $\tau^+\tau^-$ through the corresponding fermionic decays \cite{Khachatryan:2014jba,Aad:2014xzb}.

\begin{figure}[t]
\centering
\includegraphics[width=7.4cm,clip]{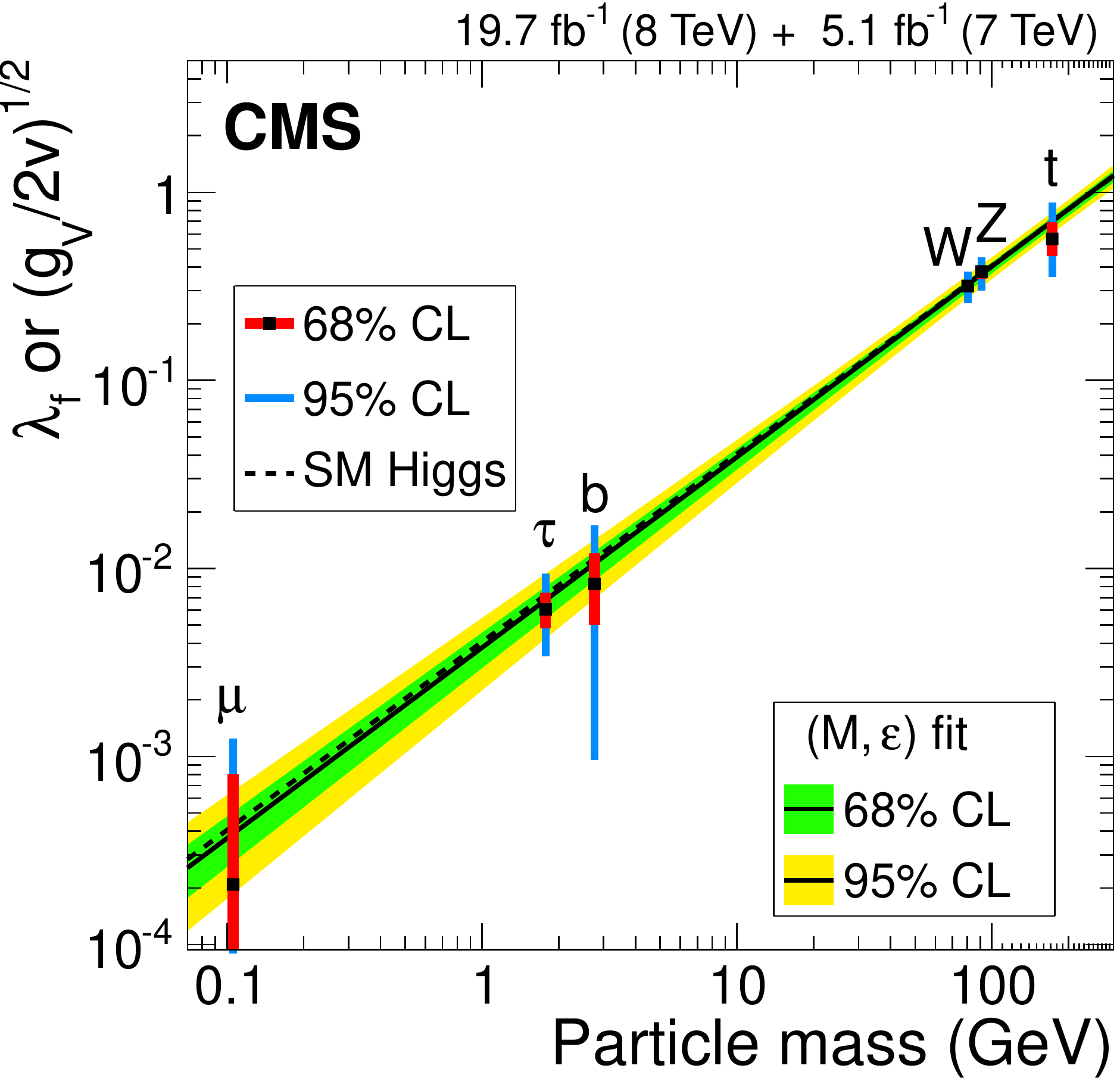} 
\caption{Higgs couplings to different particles 
\cite{Khachatryan:2014jba}.}
\label{fig:Hcouplings}
\end{figure}

The mass dependence of the measured Higgs couplings is shown in Fig.~\ref{fig:Hcouplings}. Fitting the data with the parametrization
$y_f  = \sqrt{2}\, (m_f/M)^{1+\epsilon}$\ and\
$(g_{HVV}^{\phantom{2}}/2v)^{1/2} = (M_V/M)^{1+\epsilon}$ \cite{Ellis:2013lra}
with $v = (\sqrt{2} G_F)^{-1/2} = 246$~GeV, 
one gets
$M\in [217,\, 279]$~GeV and $\epsilon\in [-0.054,\, 0.100]$ (95\% CL) \cite{Khachatryan:2014jba}, in agreement with the SM values $(M,\,\epsilon) =(v,\, 0)$. Moreover, the 95\% CL upper limit $\mathrm{Br}(H\to e^+e^-) < 1.9\times 10^{-3}$ \cite{Khachatryan:2014jba} confirms the suppression of the electronic coupling.

An important question to address is whether $H$ corresponds to the unique scalar boson incorporated in the SM, or it is just the first signal of a much richer scenario of electroweak symmetry breaking (EWSB). Obvious possibilities are an extended scalar sector with additional fields or dynamical (non-perturbative) EWSB generated by some new underlying dynamics. Whatever the answer turns out to be, the LHC finding represents a truly fundamental discovery with far reaching implications. If $H$ is an elementary scalar (the first one), one would have established the existence of a bosonic field (interaction) which is not a gauge force. If instead, it is a composite object,
a completely new underlying interaction should exist.

A fundamental scalar requires some protection mechanism to stabilize its mass. If there is new physics at some heavy scale $\Lambda_{\mathrm{NP}}$, quantum corrections could bring the scalar mass $M_H$ to the new-physics scale $\Lambda_{\mathrm{NP}}$:
\bel{eq:MH-corrections}
\delta M_H^2\,\sim\, \frac{g^2}{(4\pi)^2}\, \Lambda_{\mathrm{NP}}^2\,\log{(\Lambda_{\mathrm{NP}}^2/M_H^2)}\, .
\ee
Which symmetry keeps $M_H$ away from $\Lambda_{\mathrm{NP}}$? Fermion masses are protected by chiral symmetry, while gauge symmetry protects the gauge boson masses; those particles are massless when the symmetry becomes exact.
Supersymmetry was originally advocated to protect the Higgs mass, but according to present data this no-longer works `naturally'. Another possibility would be scale symmetry, which in the SM is broken by the Higgs mass; a naive dilaton is basically ruled out, but there could be an underlying conformal theory at $\Lambda_{\mathrm{NP}}$. Dynamical EWSB with light pseudo-Goldstone particles at low energies remains also a viable scenario.

\begin{figure}[t]
\centering
\includegraphics[width=7.55cm,clip]{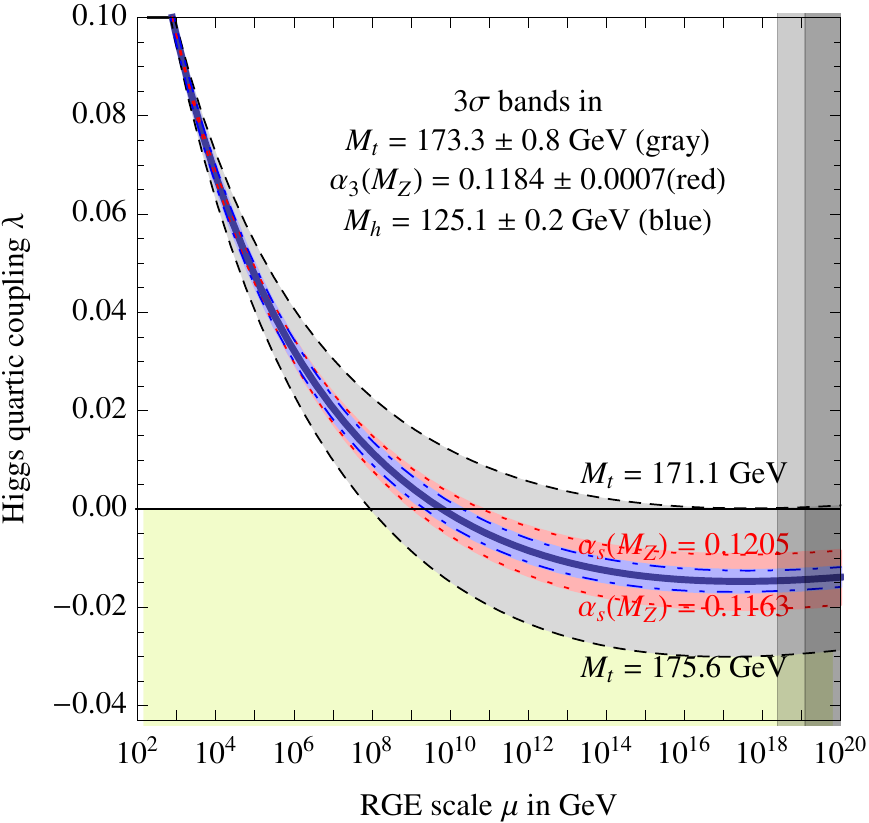} 
\caption{Evolution of $\lambda(\mu)$ with the renormalization scale  \cite{Buttazzo:2013uya}.}
\label{fig:lambda}
\end{figure}

With a Yukawa coupling very close to one,
\bel{eq:TopYukawa}
y_t\, = \,\frac{\sqrt{2}}{v}\, m_t\, =\, 2^{3/4} G_F^{1/2} m_t \,\approx\, 1\, ,
\ee
the top quark is a very sensitive probe of EWSB.
Virtual top loops dominate the electroweak quantum corrections to $M_H^2$, which grow logarithmically with the renormalization scale $\mu$:
\bel{eq:MH_QC}
\frac{M_H^2}{2 v^2} \,\approx\, \lambda(\mu) + \frac{2 y_t^2}{(4\pi)^2}\left[\lambda + 3 (y_t^2-\lambda)\,\log{(\mu/m_t)}\right] .\;
\ee
As expected, $M_H$ is brought close to the heaviest SM scale $m_t= y_t v/\sqrt{2}$.
The measured Higgs mass determines the quartic scalar coupling $\lambda$, the last free parameter of the SM. Including the positive quantum corrections, the tree-level contribution $2 v^2\lambda(\mu)$ decreases with increasing $\mu$. Fig.~\ref{fig:lambda} shows the evolution of $\lambda(\mu)$ up to the Planck scale ($M_{\mathrm{Pl}} = 1.2\times 10^{19}$~GeV), at the NNLO, varying $m_t$, $\alpha_s(M_Z^2)$ and $M_H$ by $\pm 3\sigma$ \cite{Buttazzo:2013uya}. The quartic coupling remains weak in the entire energy domain below $M_{\mathrm{Pl}}$ and crosses $\lambda=0$ at very high energies, around $10^{10}$~GeV. The values of $M_H$ and $m_t$ are very close to those needed for absolute stability of the potential ($\lambda >0$) up to  $M_{\mathrm{Pl}}$, which would require $M_H > (129.6\pm 1.5)$~GeV \cite{Buttazzo:2013uya}
($\pm 5.6$~GeV with more conservative errors on $m_t$ \cite{Alekhin:2012py}). Even if $\lambda$ becomes slightly negative at very high energies, the resulting potential instability leads to an electroweak vacuum lifetime much larger than any relevant astrophysical or cosmological scale. Thus, the measured Higgs and top masses result in a metastable vacuum \cite{Buttazzo:2013uya} and the SM could be valid up to $M_{\mathrm{Pl}}$.

\section{The Intriguing Flavour Structure}
\label{sec:Flavour}
%

The SM Higgs mechanism introduces masses in a gauge-invariant way but does not predict their values. The fermion masses and mixings are determined by the Yukawa couplings, which are arbitrary matrices in flavour space. Their diagonalization leads to a three-generation quark mixing matrix $V_{u_i d_j}$, involving three angles and one CP-violating phase. While the SM does not provide any real understanding of flavour, it accommodates all quark-flavour phenomena studied so far through this mixing matrix with a characteristic hierarchical structure.

The success of the SM description of flavour is deeply rooted in the unitarity of $V_{u_i d_j}$ and the associated GIM mechanism \cite{Glashow:1970gm}
which guarantees the absence of flavour-changing neutral currents (FCNCs) at tree level. The subtle SM cancelations suppressing FCNC transitions would be easily destroyed in the presence of new physics contributions. Therefore,
flavour data provide very strong constraints on models with additional sources of flavour symmetry breaking and probe physics at energy scales not directly accessible at accelerators.
For instance, an effective $\Delta B = 2$ interaction of the form
\bel{eq:DB=2}
\cL_{\Delta B = 2}\, =\, \frac{c^{\Delta B = 2}}{\Lambda_{\mathrm{NP}}^2}\; \left( b_L\gamma^\mu d_L\right) (b_L\gamma_\mu d_L)\, ,
\ee
induced by new physics at the scale $\Lambda_{\mathrm{NP}}$, is tightly constrained by the measured amount of $B^0$--$\bar B^0$ mixing:  $|c^{\Delta B = 2}/\Lambda_{\mathrm{NP}}^2| < 2.3\times 10^{-6}\;\mathrm{TeV}^{-2}$. A generic flavour structure with
$c^{\Delta B = 2}\sim\cO(1)$ is excluded at the TeV scale. New physics at $\Lambda_{\mathrm{NP}}\sim 1$~TeV would only become possible if $c^{\Delta B = 2}$ inherits the strong SM suppressions induced by the GIM mechanism.

\subsection{Anomalies in Rare Decays?}

The SM GIM suppression of loop-induced rare decays makes them a good testing ground for new physics. CMS and LHCb have recently measured the decays $B^0_{(s)}\to\mu^+\mu^-$ \cite{CMS:2014xfa}:
%
\beqn\label{eq:Bmm}
\mathrm{Br}(B_s^0\to\mu^+\mu^-) & = & (2.8\,{}^{+\, 0.7}_{-\, 0.6})\times 10^{-9}\, ,
\no\\
\mathrm{Br}(B^0\to\mu^+\mu^-) & = & (3.9\,{}^{+\, 1.6}_{-\, 1.4})\times 10^{-10}\, ,
\eeqn
in agreement with the SM predictions \cite{Bobeth:2013uxa} $(3.65\pm 0.23)\times 10^{-9}$ and $(1.06\pm 0.09)\times 10^{-10}$, respectively. These rates are sensitive to new physics contributions from extended scalar sectors \cite{Li:2014fea}.

Analyzing the angular distribution of $B^0\!\to\! K^{*0}\mu^+\mu^-$ decays, LHCb
\cite{Aaij:2013qta} finds a $3.7\,\sigma$ deviation from the SM \cite{Descotes-Genon:2014uoa,Altmannshofer:2014rta} for a particular optimized observable representing the interference between the longitudinal and perpendicular $K^{*0}$ amplitudes.
Another anomaly has shown up in $B^+\to K^+\ell^+\ell^-$, where the ratio between produced muons and electrons for dilepton invariant masses between 1 and 6
$\mathrm{GeV}^2$ is found to be $R_K = 0.745{}^{+0.090}_{-0.074}\pm 0.036$ \cite{Aaij:2014ora}, $2.6\,\sigma$ below the SM value~\cite{Bobeth:2011nj}.

A different flavour anomaly was found by BaBar in the tree-level decays $\bar B\to D^{(*)}\ell^-\bar\nu_\ell$ \cite{Lees:2012xj}, with a measured ratio between $\ell=\tau$ and $\ell=\mu, e$ events significantly higher than the SM prediction (2.0 and $2.7\,\sigma$ for $D$ and $D^*$) \cite{Fajfer:2012jt,Celis:2012dk}. This discrepancy has been recently confirmed by Belle \cite{Belle} and LHCb \cite{Aaij:2015yra}.
Another puzzling result is the like-sign dimuon charge asymmetry measured by D0 \cite{Abazov:2013uma}, which is $3.6\,\sigma$ above the expected SM prediction from $B^0_{d,s}$ mixing.
While more precise data is needed to clarify the situation, all these signals show the potential of flavour data to uncover new physics at higher scales.

\subsection{Violations of Lepton Flavour}

We have clear evidence that neutrinos are massive particles and there is mixing in the lepton sector. The solar, atmospheric, accelerator and reactor neutrino
data lead to a consistent pattern of oscillation parameters with
$\Delta m^2_{21} \equiv m^2_2 - m^2_1 > 0$ and two possible signs for
$\Delta m^2_{31}$: normal ($m_1\! <\! m_2\! <\! m_3$) and
inverted ($m_3\! <\! m_1\! <\! m_2$) hierarchy \cite{Agashe:2014kda}.
The main recent advance is the determination of a sizeable non-zero value of $\theta_{13}$, confirming the 3$\nu$ mixing paradigm:
$\sin^2{2\theta_{13}} = 0.084\pm 0.005$ \cite{An:2015rpe}.
The new generation of neutrino experiments should measure the CP-violating phase $\delta_{\mathrm{CP}}$ and resolve the mass hierarchy.

The non-zero neutrino masses indicate new physics beyond the SM. Singlet $\nu_R$ fields are an obvious possibility, allowing for right-handed Majorana masses of arbitrary size which violate lepton number by two units. A very large Majorana mass scale can explain the tiny size of the observed neutrino masses through the well-known see-saw mechanism \cite{Minkowski:1977sc,Mohapatra:1979ia}. 
Nevertheless, an enlarged SM with 3 light $\nu_R$ fields has also been shown to be a viable phenomenological scenario \cite{Canetti:2012kh}.

With $m_{\nu_i}\not= 0$, the leptonic charged-current interactions involve a flavour mixing matrix $V_L$. The neutrino oscillation data imply a pattern of leptonic mixings very different from the one in the quark sector, with all $V_L$ entries of similar size except for $(V_L)_{13}$ which is slightly smaller.
The smallness of neutrino masses induces a strong suppression of neutrinoless
transitions with lepton-flavour violation (LFV), which can be avoided in models with sources of LFV not related to $m_{\nu_i}$.

CMS has obtained the first bound on LFV in Higgs decays: $\mathrm{Br}(H\to\mu^\pm\tau^\mp)<1.51\%$ (95\% CL) \cite{Khachatryan:2015kon}. This improves by one order of magnitude the indirect constraints on the corresponding Higgs Yukawas from $\tau$ decays.

\section{Searching for New Physics}
\label{sec:NewPhysics}

Non-zero neutrino masses and a few (not yet significant) flavour anomalies are the only signals of new physics detected so far.
All direct searches have given negative results, pushing the new-physics scale beyond the reached sensitivity.

\subsection{Desperately Seeking SUSY}

Neither the Tevatron nor the LHC have found any convincing evidence of supersymmetry (SUSY). Strong lower bounds on the masses of SUSY partners have been set, surpassing the TeV in many cases. Moreover, the Higgs mass is heavier than what was expected to be naturally accommodated in the minimal SUSY model (MSSM).
Although recent calculations including higher-order corrections allow for slightly larger values of $M_H$ \cite{Hahn:2013ria}, the situation looks bad in the usual constrained models (CMSSM, NUHM1, NUHM2, etc.) \cite{Bechtle:2014yna}, where the 120 MSSM parameters are reduced to just a few (4 plus 1 sign in CMSSM). A global fit to the data is still possible with heavy SUSY masses ($\gtrsim 1$~TeV), but only if the muon anomalous magnetic moment is not included in the fit, since the measured value of $(g\! -\! 2)_\mu$ \cite{Pich:2013lsa} can no longer be explained \cite{Buchmueller:2014yva}.

With a larger set of 10 \cite{deVries:2015hva} or even 19--20 \cite{Cahill-Rowley:2014twa} free parameters, the Phenomenological MSSM (pMSSM) provides more flexibility and it is possible to find many parameter sets which are consistent with the data. In particular, compressing the SUSY spectrum, allows for still undetected light sparticles \cite{Cahill-Rowley:2014twa}.
Many SUSY variants (NMSSM, Split, High-Scale, Stealth, 5D, Natural, Folded, Twin, etc.) have been advocated to conform with the present experimental situation.
While some of them can be theoretically motivated (naturalness, dark matter, etc.), in most cases this is a data-driven search, looking for mechanisms to hide and avoid the strong data constraints. SUSY appears to be badly broken which questions its possible role in protecting the electroweak scale.

\subsection{Flavour in Extended Scalar Models}

The  non-generic nature of the flavour structure becomes apparent if one considers two (or more) scalar doublets $\Phi_a$, which increases the number of quark Yukawas:
%
\begin{figure}[b]
\centering
\includegraphics[width=7.5cm,clip]{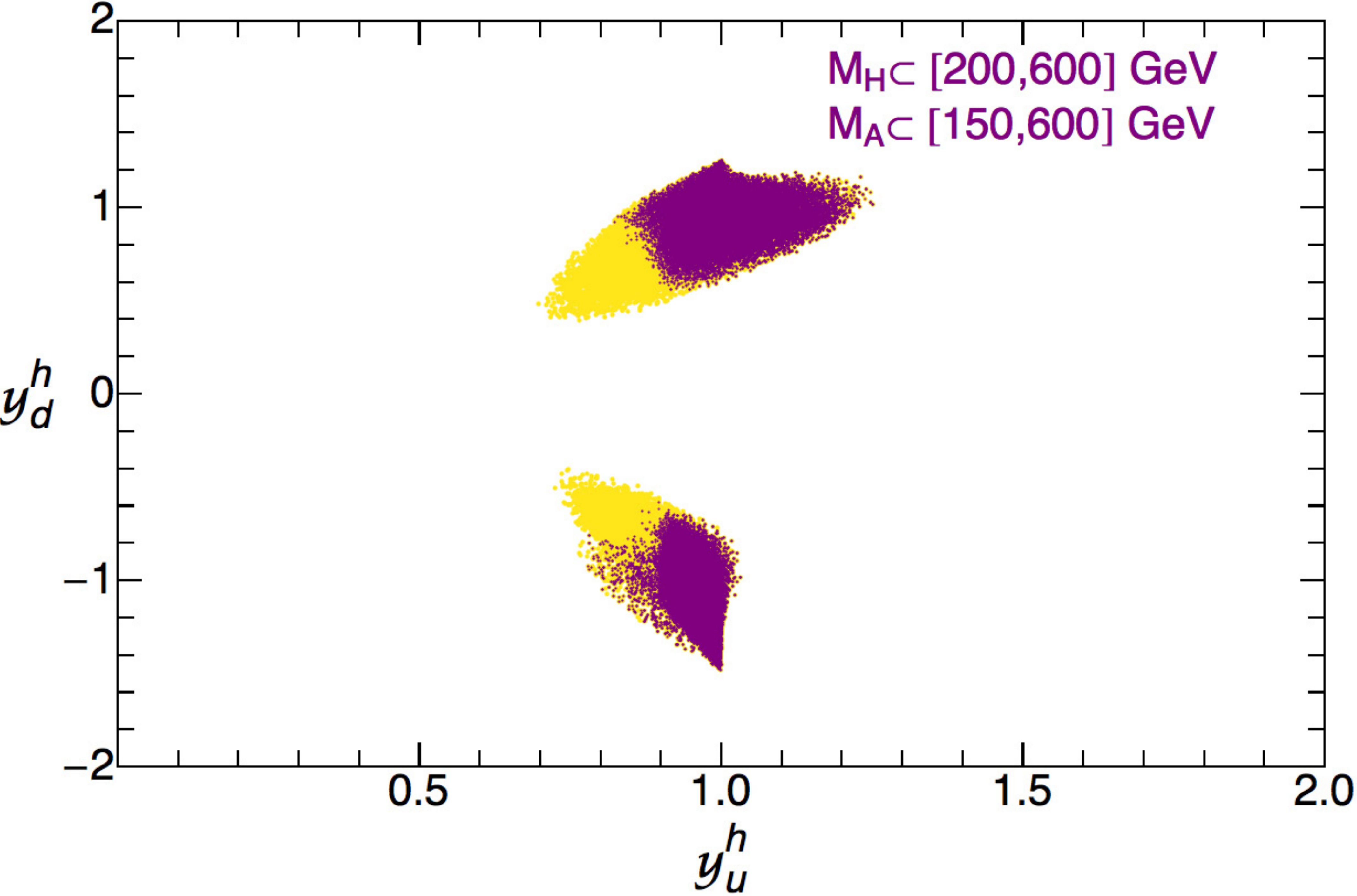} 
\caption{Allowed 90\% CL regions of the $y_q^h$ Yukawas (in SM units), in the CP-conserving A2HDM, from a fit of collider and flavour data.
The purple regions include constraints from searches of the heavier 
scalars $H$ and $A$~\cite{Celis:2013ixa}.}
\label{fig:A2HDMfit}
\end{figure}
%
\bel{eq:GenYukawa}
\cL_Y = -\sum_{a=1}^2\;\left\{ \bar Q_L'
\cY^{(a)}_d\Phi_a\, d_R'
+ \bar Q_L'\cY^{(a)}_u\tilde\Phi_a\, u_R' \right\}\, .
\ee
Here, all fermionic fields are 3-dimensional flavour vectors and
$\tilde\Phi_a \equiv i \tau_2\,{\Phi_a^*}$.
The flavour matrices $\cY^{(1)}_f$ and $\cY^{(2)}_f$ 
are in general unrelated and cannot be diagonalized simultaneously, generating dangerous FCNCs. Unless the Yukawa couplings are very small or the scalar bosons very heavy, a very specific flavour structure is required to satisfy the stringent experimental limits. The usually adopted solution imposes a discrete $Z_2$ symmetry to force one of the two Yukawa matrices to be zero \cite{Glashow:1976nt,Paschos:1976ay}; this leads to five different models with `natural flavour conservation' (types I, II, X, Y and inert) \cite{Branco:2011iw}.
A more general possibility is to require the alignment in flavour space of
$\cY^{(1)}_f$ and $\cY^{(2)}_f$ (proportional matrices),
which eliminates FCNCs at tree level \cite{Pich:2009sp,Jung:2010ik}. The three complex alignment parameters $\varsigma_f$ ($f=u,d,\ell$) introduce new sources of CP violation \cite{Jung:2013hka}.
The aligned two-Higgs doublet model (A2HDM) contains five physical scalars ($h$, $H$, $A$ and $H^\pm$), leading to a rich collider phenomenology \cite{Celis:2013ixa}.

\subsection{Looking into the Dark Side}

Several astrophysical and cosmological evidences indicate that dark matter (DM)
is the dominant matter component in our Universe, accounting for 26.8\% of its
total energy budget \cite{Ade:2013zuv}. Weakly interacting massive particles (WIMPs) around the TeV scale are considered among the leading DM candidates, because they would have the right annihilation cross section (WIMP miracle) in the early Universe, after the thermal freeze-out, to explain the present DM relic density.
Very light axion-like particles are also an alternative DM possibility.

Viable (neutral, cold and stable) DM candidates exist in many models, especially those inhibiting their decay through some symmetry, such as the lightest SUSY (R parity) or little-Higgs T-odd particles, or a $Z_2$-odd scalar in the inert two-Higgs doublet model.
The experimental bounds on DM cross sections strongly constrain the parameter space of these models.

DM could also be associated with a hidden sector, {\it i.e.} new particles that are singlets under the SM gauge group. They could be accessible through their couplings with SM singlet operators. For instance, the operator\ $\bar L_i \tilde\Phi$ could couple to new neutral singlet fermions (neutrino portal), and a new Abelian gauge field strength $F'_{\mu\nu}$ could be detected through its mixing with the SM $U(1)_Y$ field (vector portal: $F'_{\mu\nu}F^{\mu\nu}_Y$). The square of the Higgs field, $\Phi^\dagger\Phi$, provides now a very interesting scalar portal to be explored, coupling either to new singlet scalars ($S$, $S^2$) or fermion bilinears ($\bar\psi\psi$).

\section{Outlook}

Although the SM could be valid up to arbitrary high scales, new dynamics should exist because we are lacking a proper understanding of important physical phenomena, such as the matter-antimatter asymmetry, the pattern of flavour mixings and fermion masses, the nature of dark matter or the accelerated expansion of the Universe.
The SM accommodates the measured masses, but it does not explain the vastly different scales spanned by the known particles. The dynamics of flavour and the origin of CP violation are also related to the mass generation. The Higgs boson could well be a window into unknown dynamical territory, may be also related to the intriguing existence of massive dark objects in our Universe.
Therefore, the Higgs properties must be analyzed with high precision to uncover any possible deviation from the SM.

So far, the LHC did not find any exotic object and the scale of new physics has been pushed to higher energies, well above the TeV. How far this scale could be is an open question of obvious experimental relevance.
The LHC data are challenging our previous ideas about naturalness and the TeV scale.
The most fashionable new-physics scenarios are now less compelling than before,  making us suspect that Nature has chosen a quite different path.
Clearly, the LHC constraints should imply major changes in our theoretical guidelines/prejudices when searching for new-physics explanations to the many open questions that the SM leaves unanswered. The new LHC run which is just starting could bring unexpected surprises, changing our views on fundamental physics.

\section*{Acknowledgments}

I would like to thank the organizers for the invitation to present this talk.
This work has been supported by the Spanish
Government and ERDF funds from the EU Commission [Grants
No. FPA2011-23778, FPA2014-53631-C2-1-P and CSD2007-
00042 (Consolider Project CPAN)] and by Generalitat
Valenciana under Grant No. PrometeoII/2013/007.


\end{document}